# Parametric patterns in optical fiber ring nonlinear resonators

K. Staliunas[1], Chao Hang[2,3], V. V. Konotop[2,4]

[1]ICREA, Departament de Fisica i Enginyeria Nuclear, Universitat Politècnica de Catalunya,

Colom, 11, E-08222 Terrassa, Barcelona, Spain

[2]Centro de Física Teórica e Computacional, Faculdade de Ciências,

Universidade de Lisboa, Instituto para Investigação Interdisciplinar, Avenida Professor

Gama Pinto 2, Lisboa 1649-003, Portugal

[3]State Key Laboratory of Precision Spectroscopy and Department of Physics,

East China Normal University, Shanghai 200062, China

[4] Departamento de Física, Faculdade de Ciências,

Universidade de Lisboa, Campo Grande, Ed. C8, Piso 6, Lisboa 1749-016, Portugal

*Abstract.* We propose that parametrically excited patterns, also known as Faraday patterns, can be observed in nonlinear fiber resonators, where the coefficient of Kerr nonlinearity is periodically varying along the fiber in resonator. We study the parametric instability analytically on the basis of the Floquet theory, also numerically, by direct integration of the system. Instead of classical Faraday wave excitation scenario, where modulation in time causes formation of patterns in space, here we propose an inverted scenario, where the modulation in space excites the patterns in time.

*I.Introduction.*

Parametrically excited patterns, or equivalently Faraday patterns in spatially extended nonlinear systems, can emerge when a parameter of the system is periodically modulated in time, e.g. with the frequency $2\omega_0$. Then the waves (the spatial modes, or spatial harmonics) with wave-numbers centered at around a critical wave-number $k_0$ can be excited, such that their oscillation frequency are equal to the half of the frequency of parametric excitation, i.e. with $\omega_0$, related with $k_0$ via the dispersion relation of nonlinear media $\omega(k)$. Such patterns were initially observed by M. Faraday on a surface of periodically shaken (in vertical direction) mercury layer [1], where, formally, the magnitude of gravitation force is periodically modulated in time. The Faraday-like patterns were subsequently observed in vertically shaken granular media [2], in periodically driven spatially extended chemical systems [3,4], in spatially extended nonlinear optical systems [5]. Not only the extended Faraday patterns were observed [6], but also the localized structures in mechanical [7], chemical [8], and optical [9] systems. More recently the excitation of Faraday patterns of matter waves in repulsive (defocusing-type) Bose-Einstein condensates (BECs) subjected to temporal periodic modulation of nonlinearity coefficient (the atomic scattering length) was predicted theoretically [10,11], and observed experimentally [12].

A parametric pattern formation, in principle, is possible in nonlinear optics, for the light waves (either spatial beams, or temporal pulses) propagating along the matter with the Kerr nonlinearity modulated *in space*, in longitudinal (along the propagation) direction. Like in the above references [1-12], the modulationally stable case is considered, i.e. the sign of nonlinearity is such that no filamentation in two dimensional (2D) beams, and no soliton formation from 1D pulses is possible. If nonlinearity coefficient of the media is modulated along the propagation direction with the spatial wavenumber $2q$, then, simply speaking, the "flying" photon experiences a temporal modulation of nonlinearity along the propagation coordinate with the corresponding frequency $2\omega_0 = c \cdot 2q$. This sets the analogy with the parametric pattern excitation by modulating the system in time, as in previous references [1-12]. The process then leads to the transverse modulation of the light beams by $k_\perp$ (for transversally extended systems), or to longitudinal modulation of light pulses by $k_\parallel$ (in optical fibers). The selection of $k_\perp$ or $k_\parallel$ is determined by the diffraction or dispersion relations of the mater depending on the geometry (whether we consider monochromatic beams or wave pulses). The mathematical description of Faraday instability in the above presented cases in optics is identical to the Faraday instability in BECs [10,11], as the paraxial wave propagation equation for beams, or wave envelope equation for pulses are analogous to the Gross-Pitaevskii equation.

The parametric instability of the pulses propagating along the nonlinear fibers has been studied in [13] where a periodically modulated Kerr nonlinearity in normal dispersion regime was considered. A parametric excitation of transverse modulation (transverse Faraday instability) in nonlinear optics has never been considered for propagating waves to the best of our knowledge, although its mathematical description should be analogous to the longitudinal Faraday effect analyzed in [13]. In the present work we describe how one can realize very naturally the parametric Faraday instability in optical systems by using Kerr-nonlinear fiber ***resonators***.

Let us assume, that in a fiber resonator, as illustrated in Fig.1, the nonlinearity coefficient along the resonator is modulated with the spatial period equal to the half of the resonator length, i.e. that the nonlinearity makes two modulation periods along the roundtrip. In this case the longitudinal parametric instability of the light pulses could be supported by the resonator, as the period of the emerging modulation being half the period of the nonlinearity modulation would perfectly fit the length of the resonator, i.e. would be supported by the resonator condition.

Before starting the mathematical description, we inspect the proposed system in terms of the fiber resonator modes. Let a mode $k_m = 2\pi n/L$, which has the frequency $\omega_m$ following from the dispersion relation of fiber, is excited by the external injection, i.e. by the external wave with the frequency $\omega_{inj}$ close to $\omega_m$. The modulation of nonlinearity along the resonator excites the quasi-modes with $m+2$ and $m-2$ indices. The parametric excitation process is as follows: two excitations, one from the ground state mode $(k_m, \omega_m)$ and the other from the nonresonant quasi-state $(k_{m+2}, \omega_m)$ get mixed and annihilate, and two new resonant excitations appear at $(k_1, \omega_1)$ and $(k_2, \omega_2)$, obeying the energy and momentum conservation relations: $\omega_1 + \omega_2 = 2\omega_m$, $k_1 + k_2 = k_m + k_{m+2}$.

The described mechanism is illustrated in the Fig.1.b, where the dispersion is deliberately exaggerated, as in typical fiber resonators, the curvature of the dispersion curve is very small, consequently very remote modes enter into the resonance and are parametrically excited. The resonance conditions can be estimated considering the series expansion of the material dispersion $k(\omega) = k_m + \beta_1(\omega - \omega_m) + \beta_2(\omega - \omega_m)^2/2 + ...$ . The frequencies of the parametrically excited modes more conveniently can be written in the form $\omega_{1,2} = \omega_m \pm \omega_0$, where the parametrically excited sideband frequency obeys: $\omega_0 = \sqrt{2q/\beta_2}$ . For a fiber loop of one meter length ($L=1m$), for the wavelength of the order $\lambda = 10^{-6} \mu m$, and for the typical dispersion of the waveguide $\beta_2 \approx 20 \, ps^2/km$, the sideband frequency is $\omega_0 \approx 2.2 \cdot 10^{13} \, s^{-1}$, i.e. by 4 orders of magnitude larger than the mode frequency separation for the same resonator: $\Delta\omega \approx 3 \cdot 10^9 \, s^{-1}$.

A closer inspection of the mode representation in Fig.1 suggests a generalization to arbitrary number of the periods of nonlinearity along the resonator. This technically can be realized from *2p* pieces of fibers with two different nonlinear coefficients, joined in alternating order. Therefore in following, in analytic study we will consider the general case of *p*-period modulation along the resonator loop.

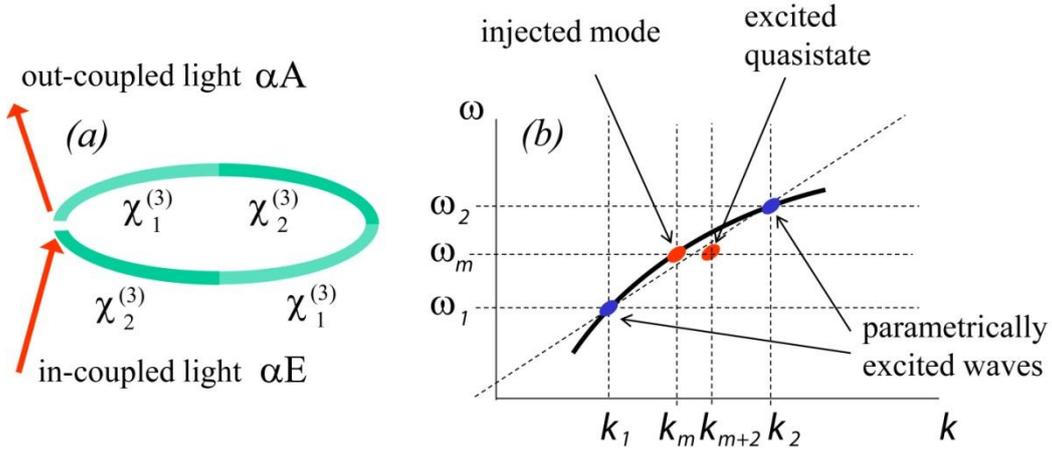

***Fig.1.*** *(Color online) a) Fiber resonator with the nonlinearity coefficient varying periodically (making two periods) along the resonator length. b) The longitudinal mode structure and the dispersion relation of the resonator. (The curvature of the dispersion curve is exaggerated.) The injected mode $(k_m, \omega_m)$ and the excited nonresonant quasi-state $(k_{m+2}, \omega_m)$ are shown by red ellipses (on horizontal dashed line at $\omega_m$); the parametrically excited modes at $(k_1, \omega_1)$ and $(k_2, \omega_2)$ are shown by blue ellipses (on horizontal dashed lines at $\omega_1$ and $\omega_2$).*

## II. The model.

We consider the light propagation along the fiber by solving the nonlinear Schrödinger equation (NLSE) for the field envelope $A^{(n)}(\tau, z)$:

$$\partial A^{(n)}(\tau,z)/\partial z = -i\, \partial^2 A^{(n)}(\tau,z)/\partial \tau^2 + ic(z)\left|A^{(n)}(\tau,z)\right|^2 A^{(n)}(\tau,z) \qquad (1.a)$$

defined for each (the *n*-th) resonator roundtrip; respectively $\tau$ varies between 0 and T, where T is the period of a round trip. The reference frame moves with the group velocity of light, therefore the first time derivatives of the field envelopes do not appear in (1.a). The retarded time $\tau$ is scaled to make the second order dispersion coefficient unity. The nonlinearity coefficient is considered periodic, with the period $L/(2p)$, where $p$ is an integer and $L$ is the length of the resonator. The modulation wavenumber is then $2q = 4\pi p/L$. In numerical calculations we consider harmonic modulation of the coefficient of nonlinearity $c(z) = c_0 + 2\Delta c \cdot \cos(2qz)$, however the basic results hold for arbitrary (*e.g.* a step-like) function of nonlinearity. Important is that the average nonlinearity $c_0$ is positive. We also take care that the nonlinearity remains positive during the entire roundtrip: $2\Delta c < c_0$ in order to exclude any possibility that the system enters into modulationally unstable regime during a single round trip (i.e. the system remains all the time in the modulationally stable regime).

Eq. (1.a) is to be completed by the condition relating the fields at the end ($z = L$) of the current (*n*-th) resonator roundtrip, and at the beginning ($z = 0$) of the next (*n+1*) roundtrip:

$$A^{(n+1)}(\tau,0) = (1-\alpha)A^{(n)}(\tau,L)e^{i\varphi} + \alpha E(\tau) \tag{1.b}$$

which means that: i) an external field is injected with the envelope $E(\tau)$, which is either a continuous wave $E(\tau) = const = E_0$, or a pulse from the pulse train of period synchronized with the length of the resonator; ii) a fraction $\alpha$ of the field is decoupled from the fiber resonator; iii) the propagation along the fiber results in some linear phase shift $\varphi$ with respect to the external injection, which is either due to particular reference frame of NLSE (1.a), or due to out-coupling mechanism introducing a particular phase shift.

The system (1) is a well-known Ikeda map [14,15] widely used to simulate light propagation in fiber lasers and resonators. In particular Eqs. (1) were used in [16,17] to study the stability of Kerr-resonator without any spatial modulation of nonlinearity. We note however, that the system (1) does not require the continuity and differentiability of the field at the ends of the resonator, i.e. generally speaking $A^{(n+1)}(\tau,0) \neq A^{(n)}(\tau,L)$ (unless a specific form of the solution is chosen, see (2) below, or the pulsed injection case is considered). Moreover, it was shown in [18] that the continuity condition can eliminates some instabilities of the system, i.e. in other words some instabilities observed in [16,17] can occur due to the above mentioned discontinuity. In this context we also mention that alternatively to Ikeda map (1), which is integrated throughout the present paper, continuous wave model are being developed [18,19], where the field persists continuously when passing through the cavity input (output). The parametric instability can be also demonstrated in continuous wave models as our preliminary results indicate, here, however, we restrict to analysis using a more simple Ikeda model (1).

### III. Stability analysis.

**III.1. Homogeneous solution of the map.** We start with the linear stability analysis of a homogeneous solution, i.e. we assume that the injection is a continuous wave: $E(\tau) = const = E_0$. The respective solution of (1.a) reads:

$$A^{(n)}(z) = A_0^{(n)} e^{i|A_0^{(n)}|^2 \int c(z)dz} \tag{2}$$

At the end of each resonator roundtrip: $A^{(n)}(L) = A_0^{(n)} e^{i|A_0^{(n)}|^2 c_0 L} = A_0^{(n)} e^{i\theta}$, where $A_0^{(n)}(0) = A_0$ is a constant field and $\theta = |A_0^{(n)}|^2 c_0 L$ is a nonlinear phase shift accumulated in propagation along the fiber. The resonator end-roundtrip condition (1.b) in stationary case allows calculating the amplitude of the field as the solution of transcendental equation:

$$A_0 = (1-\alpha)A_0 e^{i\varphi+i\theta} + \alpha E_0 \qquad (3)$$

***III.2. Linear stability analysis.*** Next we perform linear stability analysis of the map (1) with respect to the sideband modes. We consider in a standard way the exponential-oscillatory growth of a small harmonic mode of perturbation:

$$A^{(n)}(z,\tau) = A_0^{(n)}(z)\left(1 + a^{(n)}(z)\cos(\omega\tau)\right) \qquad (4)$$

where $\omega = \dfrac{2\pi}{T}$. The resonator end-roundtrip condition also applies to the perturbation:

$$a^{(n+1)}(0) = (1-\alpha)a^{(n)}(L)e^{i(\phi+\theta)} \qquad (5)$$

as it follows from (1.b) and (3). We also notice that the ansatz (4) implies that:

$$A^{(n)}(z,0) = A^{(n)}(z, 2\pi/\omega) \qquad (6)$$

what can be viewed as a constraint (boundary conditions on the domain $[0,T]$) additional to (1.b).

A goal of the analysis is the calculation of the Floquet multipliers (the multipliers indicating the growth of perturbation on one full resonator roundtrip) depending on the frequency of the emerging modulation $\omega$. If the modulus of the largest multiplier is larger than 1, then the map is unstable, and the corresponding modulation mode grows.

It is known, that even the stability analysis of the NLSE with periodically modulated nonlinearity cannot be performed analytically without simplifications [20,21]. Here we face additionally the effects of the resonator end-roundtrip condition which makes the analysis even more complicated. Therefore we follow two alternative approaches: i) numerical approach, by calculating the evolution of perturbations numerically over full resonator roundtrip (the propagation along the nonlinear fiber plus the resonator end condition, i.e. losses, phase shift, and the injection), and then calculating the Floquet multipliers numerically by diagonalizing the perturbation evolution matrix; ii) simplified analytic approach, considering a weak modulation of nonlinearity and a weak and harmonic response.

***III.3. Numerical stability analysis.*** A standard technique is used: we find the homogeneous stationary solution (3), add a small harmonic perturbation $a^{(n)} = \varepsilon\cos(\omega\tau)$, and calculate the evolution of the perturbation over a full resonator roundtrip integrating numerically (1). We ensure that the perturbation is sufficiently small (we are still in a linear regime with respect to the perturbation) checking that the modulation at the other harmonics of the excitation do not appear in one roundtrip. The procedure is repeated for a set of two independent perturbations $a_1^{(n)} = \varepsilon\cos(\omega\tau)$ and $a_2^{(n)} = i\varepsilon\cos(\omega\tau)$. In this way the map is numerically calculated and a 2x2 matrix is obtained relating the real and imaginary parts of perturbation at the beginning of n-th and (n+1)-th resonator roundtrips. The matrix is numerically diagonalized, and the Floquet multipliers are calculated, as the eigenvalues of the map. The modulus of the largest eigenvalue indicates stability of the perturbation.

Fig.2. shows the instable areas of the map. In Fig.2.a a nearly-resonant case is considered, i.e. the situation where the nonlinear phase shift of homogeneous solution $\theta = \left|A_0^{(n)}\right|^2 c_0 L$ is approximately compensated by the linear phase shift: $\theta + \varphi \approx 0$. In Fig.2.b a strongly nonresonant case is calculated. In both cases the instability areas are visible (the so called "Arnold tongues" [20,21]. The instability areas, as expected, increas with the amplitude of modulation of nonlinearity. In the close-to-resonance case the instabilities are generally stronger than in far-from-resonant case, as the amplitude of the homogeneous solution (and consequently all the nonlinear effects), is higher.

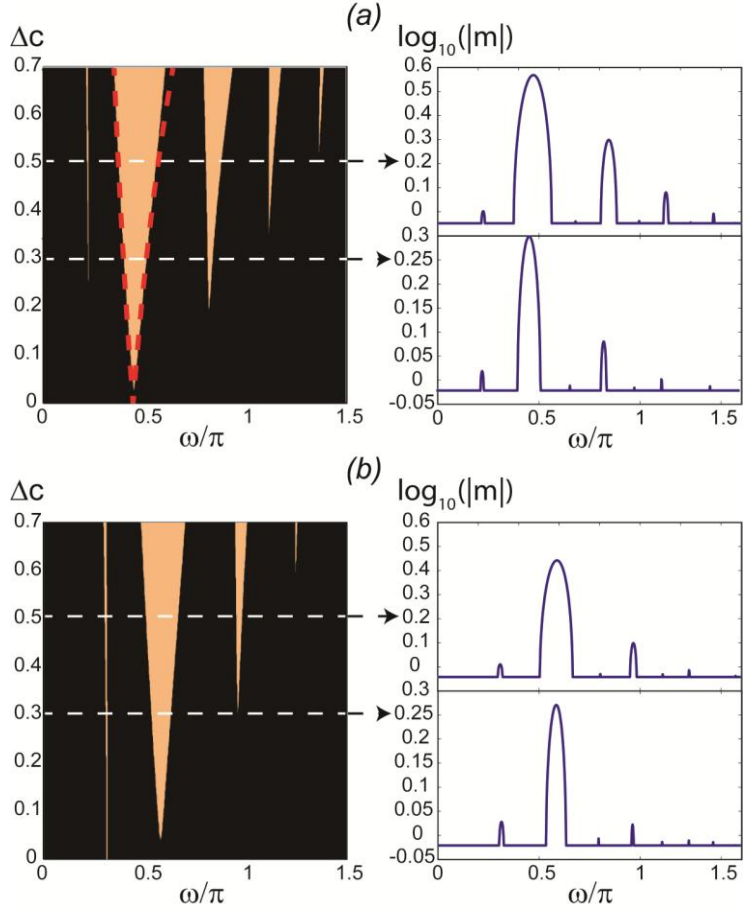

**Fig.2** *(Color online) Numerically calculated instability domains (on the plane $(\omega, \Delta c)$) for the near-resonant (a) and far-off-resonant (b) cases. The dashed lines in (a) indicate analytically calculated instability boundary corresponding to first Arnold tongue (see Eqs. (10)). Parameters: $c_0 = 1.0$, $L = 1.0$, $\alpha = 0.1$, $E_0 = 3.3$; also $\varphi = -8.5$ for (a), and $\varphi = -4.0$ for (b). The insets show the moduli of dominating Floquet multipliers in logarithmic scale ($\log_{10}(|m|)$) for different $\Delta c$: $\Delta c = 0.3$ and $\Delta c = 0.5$ (indicated by horizontal lines).*

The structure of the Arnold tongues resembles that in a dissipative NLSE, as follows e.g. from comparison with [10]. However some differences with the stability diagram of the dissipative NLSE in [10] is envisaged: Apart from the parametric instability tongues, another instability related due to presence of the resonator appears (see Fig.2.b). This instability typically appears in a far-from-resonant case, and is not related with the modulation of nonlinearity. Such instability of fiber ring resonators modeled by (1) in the normal dispersion regime has been found and discussed by Halterman et al. in [16,17].

Note also the frequency of the most unstable mode (the tip of the first Arnold tongue) depends on the linear phase shift $\varphi$. This is due to the dependence of the amplitude of stationary state $A_0$ on the linear phase shift, as follows from the transcendental equation (3).

*III.4. Approximate stability analysis.* We split the analysis into two stages: 1) the conservative nonlinear propagation along the fiber governed by modulated NLSE (1.a); 2) the transformation of fields at the end of resonator roundtrip (1.b).

1) Assuming, that the modulation is weak, the perturbation of the NLSE solution is searched in the form of expansion into spatial harmonics with the period of multiples of half of the period of modulation of nonlinearity:

$$A^{(n)}(z,\tau) = A_0^{(n)}(z)\left(1 + \left(a_0^{(n)}(z) + a_{+1}^{(n)}(z)e^{iqz} + a_{-1}^{(n)}(z)e^{-iqz} + ...\right)\cos(\omega\tau)\right) \quad (7)$$

where $q = 2\pi p/L$ is the half of the modulation wavenumber, $A_0^{(n)}(z)$ is the solution of the map in stationary regime (3), and $a_j^{(n)}(z)$ are the slowly varying amplitudes of the corresponding harmonics of the perturbation). For the sake of simplicity, below we drop the upper index (i.e. the number of round trip) since the consideration is limited to one trip only.

The expansion (7) is plugged into (1.a), and an infinite set of coupled equations for the harmonics of perturbations (in terms of $q$) is obtained. After diagonalization of the linear system the eigenvalues can be in principle calculated and, in this way, the linearized version of NLSE (1.a) can be integrated over a *2p* modulation periods (over one resonator roundtrip). This can be done only numerically, by truncating the infinite chain of equations. In order to estimate roughly the instability, a substantial simplifications is possible, by truncation of the expansion to only two harmonics, those with $q$ and $-q$ in expansion (7) (i.e. the use of only harmonic in space perturbation at a half of excitation frequency):

$$da_1/dz = i(\omega^2 - q)a_1 + ic_0|A_0|^2(a_1 + a_{-1}^*) + i\Delta c|A_0|^2(a_{-1} + a_1^*) \quad (8.a)$$

$$da_{-1}/dz = i(\omega^2 + q)a_{-1} + ic_0|A_0|^2(a_{-1} + a_1^*) + i\Delta c|A_0|^2(a_1 + a_{-1}^*) \quad (8.b)$$

The growth exponents of (8) are found by calculating eigenvalues of the matrix of the coefficients (note that the 4x4 matrix is to be diagonalized, due to complex character of (8)): The dominating eigenvalue (the one with the largest increment):

$$\lambda = \sqrt{-q^2 - \omega^2\left(2c_0|A_0|^2 + \omega^2\right) + 2\omega\sqrt{\left(\Delta c|A_0|^2\right)^2\omega^2 + q^2\left(2c_0|A_0|^2 + \omega^2\right)}} \quad (9)$$

describes the first Arnold tongue of the instability of the NLSE. As only the first harmonics is considered in the expansion (7), only the first Arnold tongue is obtained (a truncation to higher harmonics would reproduce the higher order instability tongues).

The first Arnold tongue, according to (7) is centered at the frequency:

$$\omega_0^2 = -c_0|A_0|^2 + \sqrt{\left(c_0|A_0|^2\right)^2 + q^2} \quad (10)$$

And the instability is bounded by:

$$\Delta c |A_0|^2 = \frac{q^2}{2\omega^2} - c_0 |A_0|^2 - \frac{\omega^2}{2} \quad \text{for} \quad \omega < \omega_0 \quad (11.a)$$

$$\Delta c |A_0|^2 = -\left(\frac{q^2}{2\omega^2} - c_0 |A_0|^2 - \frac{\omega^2}{2}\right) \quad \text{for} \quad \omega > \omega_0 \quad (11.b)$$

At the resonance frequency $\omega = \omega_0$ and for weak modulation: $\Delta c |A_0|^2 \ll 1$ (i.e. at around the tip of the first Arnold tongue in Fig.3), Eq. (9) simplifies:

$$\lambda = \frac{\Delta c |A_0|^2}{q} \sqrt{2(c_0 |A_0|^2)^2 + q^2 - 2c_0 |A_0|^2 \sqrt{(c_0 |A_0|^2)^2 + q^2}} \quad (12)$$

Generally, the NLSE is to be solved for the propagation along the fiber, and the transformation relating the evolution of the perturbations in a roundtrip (e.g. in the basis of $(a_1, a_1^*, a_{-1}, a_{-1}^*)$) is to be calculated as a 4x4 matrix $\hat{N}$ (not defined explicitly in the text, due its clumsy form).

2) The condition at the end of the roundtrip in the basis of $(a_1, a_1^*, a_{-1}, a_{-1}^*)$ reads (as directly follows from (5)):

$$\hat{R} = (1-\alpha)\hat{D}\left(e^{i(\varphi+\theta)}, e^{-i(\varphi+\theta)}, e^{-i(\varphi+\theta)}, e^{-i(\varphi+\theta)}\right) \quad (13)$$

Where $\hat{D}$ is a diagonal matrix.

The maps describing evolution of perturbation in nonlinear propagation $\hat{N}$, and at the end of resonator roundtrip condition $\hat{R}$ are to be considered together by calculating the matrix product $\hat{R}\hat{N}$, and diagonalized in order to obtain the Floquet multipliers for the full resonator roundtrip. Again the explicit analytical expression is impossible in general case, as the stationary values of $|A_0|^2$, and consequently the $\theta$, are given by transcendental equations (3). However, restricting to a special case of the resonance $\theta + \varphi = 0$, the resonator end round-trip matrix becomes proportional to the unit matrix $\hat{R} = (1-\alpha)\hat{I}$, and the analysis substantially simplifies. In particular the four Floquet multipliers depend solely on their "own" Lyapunov exponents of the NLSE perturbation (the NLSE perturbations modes do not couple mutually at the end of the resonator roundtrip), and read: $m_i = (1-\alpha)e^{\lambda_i L}$. Then remains to consider only the dominating instability calculated in (9), (12), and the stability condition becomes: $\lambda L + \ln(1-\alpha) > 0$, which for $\alpha \ll 1$ simplifies to $\lambda L/\alpha > 1$.

As a rough approximation: for the resonant excitation $\omega \approx \omega_0$ the largest growth exponent in conservative propagation is: $\lambda \approx \Delta c |A_0|^2$ (here (12) was simplified in short wavelength limit $q \gg (c_0 |A_0|^2)$), therefore the instability condition is: $L\Delta c |A_0|^2 > \alpha$. This is a good guiding point in search for the parametric instability in numerics and experiments. The physical interpretation of the latter relation is the following: the parametric instability in the ring Kerr-nonlinear cavity develops if the overall variation of the modulus of the nonlinear phase shift is larger than the losses during the roundtrip. In other words: if the Kerr cavity is constructed from the pieces of fibers with nonlinearity coefficients $c_1$ and $c_2$ then the characteristic nonlinear phase shift difference $L|\Delta c||A_0|^2 \approx L|c_1 - c_2||A_0|^2/2$ in a roundtrip must be of the order, or larger, than the roundtrip loss for the development of the parametric instability.

## IV. Numerical study of evolution.

We have performed full numerical simulations of the map (1) by solving consequently the NLSE (1.a) for wave propagation along the loop, then applying the resonator end-roundtrip condition (1.b). We consider two cases: i) a plane wave $E(\tau) = E_0$ is applied; ii) excitation by Gaussian pulses $E(\tau) = E_0 \exp(-(\tau - T/2)^2 / \tau_0^2)$.

The first case (the plane wave $E(\tau) = E_0$) brings to numerical demonstration of the above found and analyzed instability. In the numerical calculation, we first calculate the stationary solution from the transcendental equation (3). Then, a small $\delta$-correlated noise (typically of amplitude $10^{-3}$) is added to the stationary solution in order to trigger the instability. Finally, we solve the map (1) for 400 roundtrips and represent the field in time and frequency domain. In order to obtain a more regular evolution we stay close at the instability boundary in parameter space: we take a relatively small modulation of nonlinearity, $\Delta c = 0.175$, while keep the condition $L\Delta c |A_0|^2 > \alpha$ being satisfied. The evolution is summarized in Fig.3. The threshold and frequencies of excited modes of the instability correspond well with those estimated from the linear stability analysis.

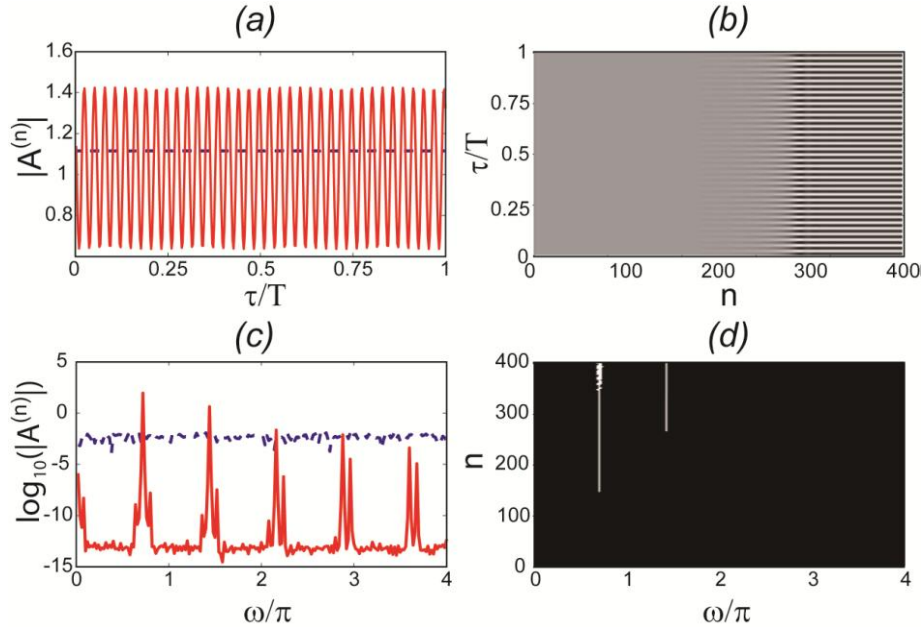

***Fig.3.*** *(color online) The field dynamics in case of constant injection $E(\tau) = E_0$. a): the envelopes (amplitudes) of the field $A^{(n)}$ for n=0 ($|A^{(0)}| = |A_0| = 1.12$; dashed line) and for n=400 (solid line). n is the number of the roundtrips. b): the evolution of the envelopes. c): the spectra in log scale ($\log_{10}(|A^{(n)}|)$) for n=0 (dashed line) and for n=400 (solid line). d): the evolution of the spectrum. The first Arnold tongue at $\omega/\pi \approx 0.7$ corresponds well with that estimated from the linear stability analysis. The parameters are: $\alpha = 0.1$, $\varphi = -1.1$, $E_0 = 2.0$, $c_0 = 1.0$, $\Delta c = 0.175$, and $L = 1.0$. A small perturbation of amplitude $10^{-3}$ is added to trigger the instability.*

For larger amplitude of modulation of nonlinearity coefficient the parametrically excited pattern was no more stationary. In particular for $\Delta c = 0.25$, periodic revivals of the emerging patterns were observed, similarly to those reported in [22] for pure parametrically modulated NLSE. For even larger amplitude of nonlinearity modulation, e.g. for $\Delta c = 0.35$, a chaotic evolution of the field as initiated by the predicted instability is obtained as shown in Fig.4.

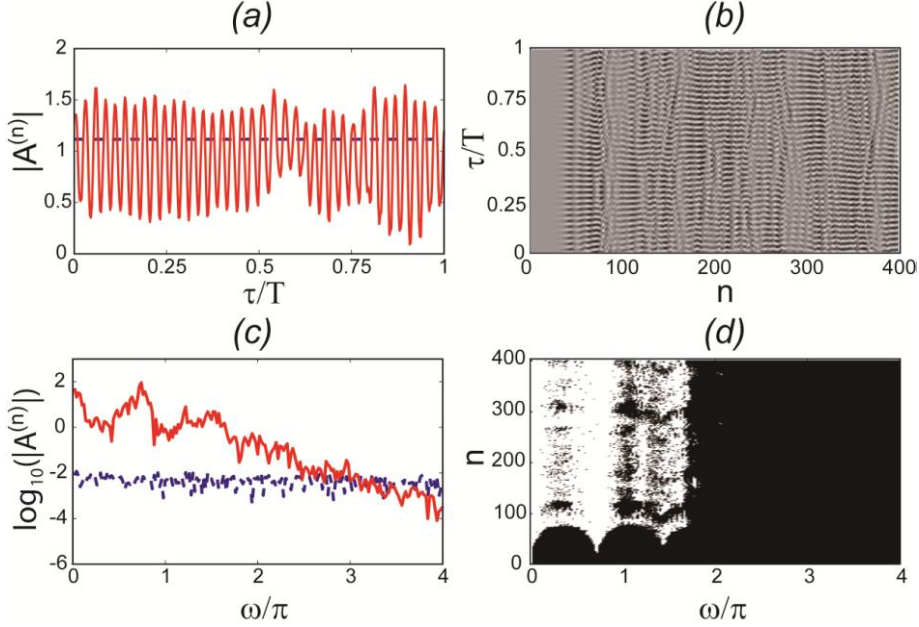

***Fig.4.*** *(color online) The spatio-temporally chaotic evolution of the field in case of constant injection $E(\tau) = E_0$. Each panel corresponds to the one with the same label of Fig. 3. The parameters are the same with those given in Fig. 3, except for $\Delta c = 0.35$.*

The second case (the excitation by Gaussian pulse $E(\tau) = E_0 \exp(-(\tau - T/2)^2 / \tau_0^2)$ with $\tau_0 = T/6$ in this particular case) shows the persistence of the instability and of the pattern formation. In the numerical calculation, we first solve the map (1) for 100 resonator roundtrips without the modulation of nonlinearity. After the stationary pulses sets in the resonator (with the envelopes different from the envelope of Gaussian pulse of injection), the modulation of the nonlinearity is switched on. In the latter case of pulsed injection, the random perturbation is not necessary to trigger the instability, differently from the previous case of homogeneous pump. The subsequent iteration for 400 roundtrips of the map (1) with the modulation of the nonlinearity indicates the growth of instable modes, as summarized in Fig. 5. The frequency of the instability in this latter case is in a good correspondence with the former case and the analytical estimation.

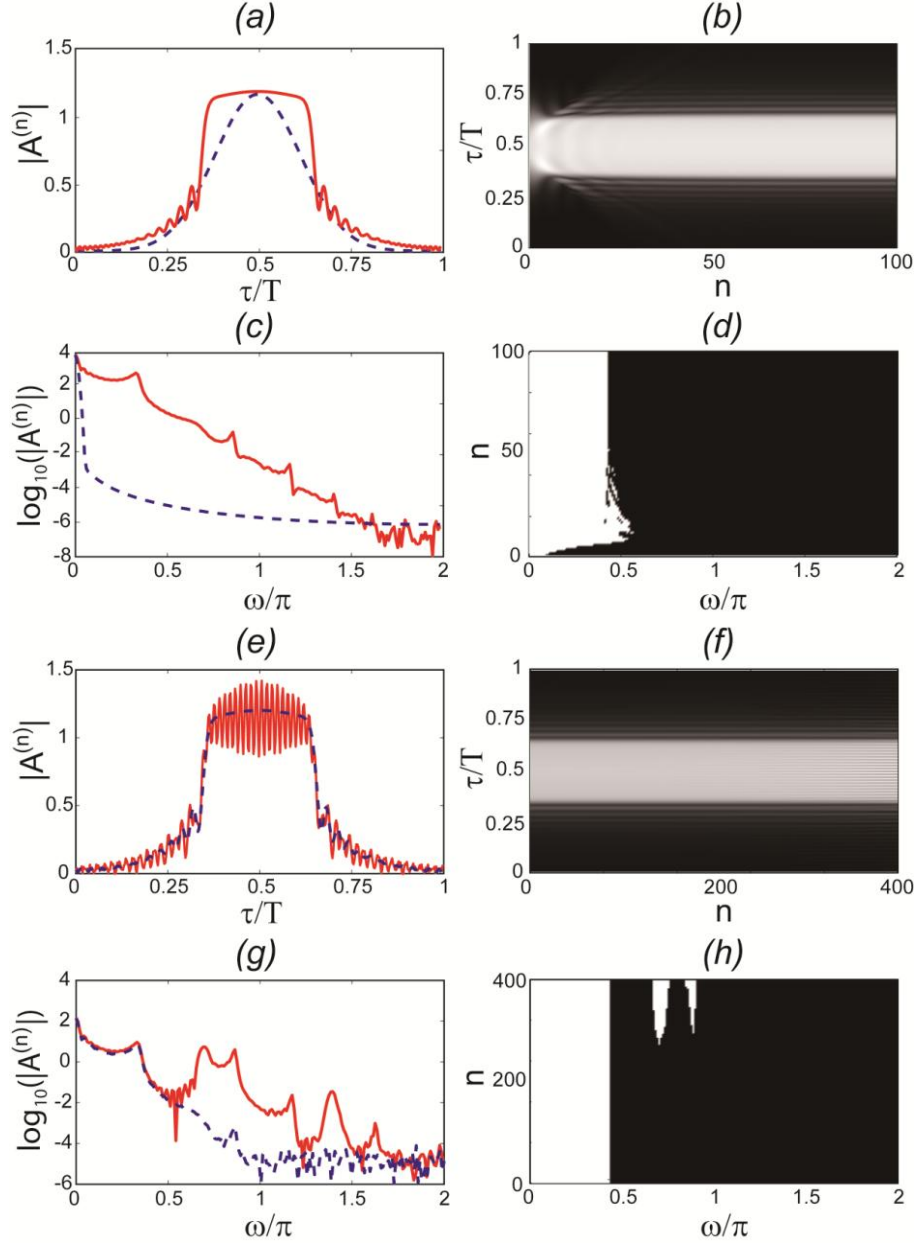

*Fig.5.* Evolution of the field for injection in form of Gaussian pulses $E(\tau) = E_0 \exp\left(-(\tau - T/2)^2/\tau_0^2\right)$ with $\tau_0 = T/6$. The first four panels, (a)-(d), are obtained with $\Delta c = 0$, while the following four panels (e)-(h), are obtained with $\Delta c = 0.206$. (a), (e): the modules of the field $A^{(n)}$ for n=0 (dashed line) and for n=100 and 400 (solid line). (b), (f): the evolution of the envelope. (c), (g): the spectra in log scale ($\log_{10}\left(|A^{(n)}|\right)$) for n=0 (dashed line) and for n=100 and 400 (solid line). (d), (h): the evolution of the spectrum. The first Arnold tongue is still at $\omega/\pi \approx 0.7$ (see panel (h)) which matches well with the estimate obtained from the linear stability analysis. The parameters are: $\alpha = 0.1$, $\varphi = -1.2$, $E_0 = 3.0$, $c_0 = 1.0$, and $L = 1.0$.

## V. Conclusions.

Summarizing, we predict the parametric instability in a Kerr-nonlinear fiber cavity. We discuss the analogy between the Kerr-nonlinear fiber resonator with the "conventional" systems showing the Faraday instability. We show the emergence of the patterns by the linear stability analysis (based on the Floquet theory). Finally we prove the pattern formation by solving the nonlinear Ikeda map.

We emphasize that the reported instability has nothing to do with the modulational instability. The latter occurs in fibers *(and fiber resonators)* in anomalous dispersion case, when also the soliton formation is possible. Here we have an opposite situation. Moreover, as the modulation instability is a long-wave instability, and here we observe the well controllable, short-wave instability, typical for parametrically excited instabilities.

Finally, we note that the character of the Faraday pattern formation process is reversed: if in the classical situation the temporal modulation of some parameters (e.g. nonlinearity) excites parametrically the spatial Faraday waves and patterns, here the spatial modulation (along the fiber) of nonlinearity excites the periodic in time modulation of wave or of the envelope of the pulse.

*Acknowledgements*. The work is financially supported by Spanish Ministerio de Educación y Ciencia and European FEDER through project FIS2011-29734-C02-01, by the FCT under the grant PEst-OE/FIS/UI0618/2011, and actiones integradas Spain-Portugal PT2009-0089 (N ºE-27/10).